\documentclass{ws-ijmpa}
\usepackage{amssymb}
\usepackage{graphicx}
\begin{document}
\title{CHARMONIA FROM LATTICE QCD}

\author{GUNNAR S. BALI\footnote{Invited talk presented at ``Charm 2006'', Beijing, 5--7 June 2006.}
}

\address{Institut f\"ur Theoretische Physik, Universit\"at Regensburg,
Universit\"atsstr.\ 32\\
93053 Regensburg, Germany\\gunnar.bali@physik.uni-r.de}

\maketitle

\begin{abstract}
Recent lattice QCD results on
charmonium properties are reviewed. I comment on
molecules and hybrid states as well as on
future studies of states near strong decay thresholds.
\end{abstract}
\ccode{PACS numbers: 12.38.Gc, 14.40.Gx, 12.39.Pn, 13.20.Gd}
\section{New States}
Recently, for the first time in over twenty years,
several narrow hadronic resonances have been discovered.
Examples are an $\Upsilon$ $D$-wave, the $B_c$,
the $\eta_c'$, the $h_c$ and also new charmed baryons.
Some findings were particularly embarrassing for theorists,
such as the $D_{s0}^*(2317)$, the
$D_{s1}^*(2460)$ and the $X(3872)$, whose masses and
properties were not anticipated. Of course, in retrospect
everyone can explain anything. The power of post diction
is symptomatic for our lack
of understanding of strong interaction dynamics.
This is further illustrated by the now
fading pentaquark-hype. What are the
relevant degrees of freedom for the classification of
hadronic excitations? Are angular and radial excitations of quark
model states sufficient to obtain a qualitative picture
or do we have to include hybrid- and
molecules (i.e. meson-meson and/or tetraquark states) as well?
(How) can we make
quantitative predictions of masses and strong decay rates?

Charmonia offer an exciting arena for improving theoretical
tools and experimental techniques. In some aspects
charm quarks can be classed as ``heavy''. Nonetheless, with
relative valence quark velocities $v>1/2$
and binding energies
$\overline{\Lambda}\gtrsim m_c/3$, it is
clear that corrections to an effective heavy quark
field theory description can be large:
ultimately there is no short-cut to the inclusion of
non-perturbative low energy QCD effects. What we
learn from charmonia will to a large extent also apply
to light hadron spectroscopy.
However, (potential) non-relativistic QCD ((p)NRQCD)
expectations can still serve as
a starting point. Also potential models with phenomenologically
fitted parameters appear to describe many features
reasonably well and it needs to be clarified to what extent
these are connected to QCD as the fundamental theory.

\begin{figure}[t]
\centerline{\includegraphics[width=0.9\textwidth,clip]{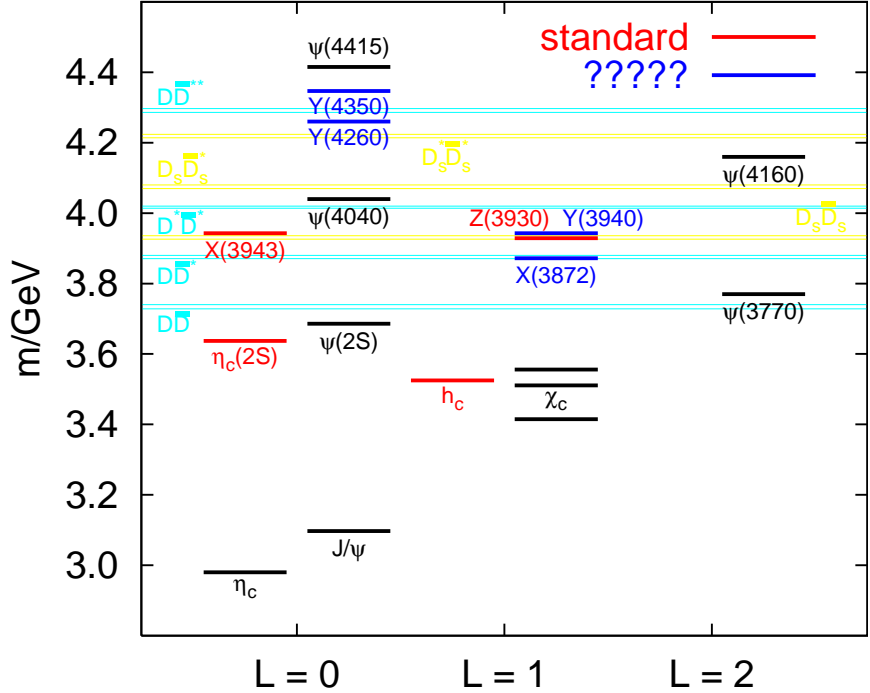}}
\caption{The eight recently discovered charmonium states.
Some can be interpreted as $c\bar{c}$ quark model states (standard)
while others do not appear to fit into the expected
pattern (?????).}\label{fig:spec}\vskip -.5cm
\end{figure}
Due to the spatial compactness of charmonium wavefunctions
we encounter the luxury of finding many narrow resonances below strong
decay thresholds. Also above such thresholds the
situation is still much cleaner than it is in the light quark sector.
While this applies even more so to bottomonia, in the $\Upsilon$
case, production in $e^+e^-$ machines is limited to vector
states and their decay products. Not even
the $\eta_b$ meson has been detected thus far! Furthermore,
the $B\overline{B}$ threshold region is elusive.
Some information could be gained by scans above
$\Upsilon(4S)$ and runs at $\Upsilon(10860)|\Upsilon(11020)$
in a future super-$B$ factory. Nonetheless, the prospects in
$\Upsilon$ spectroscopy are quite bleak due to the
small hadronic production cross sections
and the unavailability of a top-meson that could produce $\Upsilon$
states in its decay. In the charm-sector however
many new states have been discovered recently,
some in $p\bar{p}$ collisions at FNAL, others in
$\psi$ production and decays at CLEO and many by the $B$ factories,
mostly in $B$ decays. A summary of these recent findings is presented in
Fig.~\ref{fig:spec}. 10 resonances have been discovered
1974 -- 1977, none in 1978 -- 2001 and eight (!) since 2002.
For recent reviews of the properties of these new states, see e.g.\
Ref.\cite{Godfrey:2006pd}

\section{Spectroscopy: $n_f=0$ and $n_f>0$}
\begin{figure}[t]
\centerline{\rotatebox{270}{\includegraphics[height=0.9\textwidth,clip]{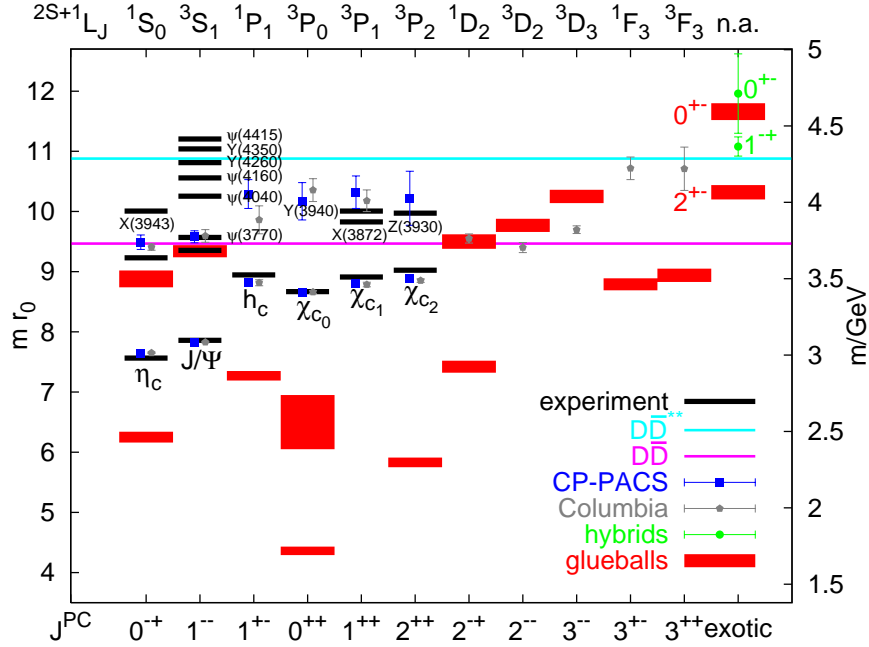}}}
\caption{The quenched charmonium spectrum
(Columbia\protect\cite{Liao:2002rj}, CP-PACS\protect\cite{Okamoto:2001jb}),
glueballs\protect\cite{Chen:2005mg}
and spin-exotic
$c\bar{c}$-glue hybrids\protect\cite{Liao:2002rj}, overlayed with the
experimental spectrum.}
\label{fig:chquench}
\end{figure}
Most hadronic resonances undergo strong decays.
This situation is hard to study on the lattice and requires
repeating the simulation on different (and large)
physical volumes. Fortunately, most charmonia are very narrow.
The states above
3900 MeV however have widths ranging
from $O(10)$~MeV to $O(100)$~MeV.
A model with strong decays switched off can serve as a starting point 
for investigations of the more involved realistic situation.
This can be achieved by switching sea quark
effects off completely.
Studies in this so-called quenched approximation
have been performed by a variety of collaborations,
with an anisotropy
parameter\cite{Liao:2002rj,Okamoto:2001jb,Mei:2002ip,Luo:2005zg,Dudek:2006ej,Dudek}
$\xi=a_s/a_t> 1$ and
keeping spatial and temporal lattice resolution the
same,\cite{Bernard:1997qe,Boyle:1999dx,Choe:2003wx,McNeile:2004wu,deForcrand:2004ia,Tamhankar:2005zi} $a_s=a_t$.

For charmonia,
$\langle r^{-1}\rangle\gg M$, where the mass $M$ is calculated in
units of $a_t^{-1}$ and the wavefunction extent $r$ in units of $a_s$.
Therefore, introducing an
anisotropy appears to allows for a smaller number of lattice points and
hence a reduction in the computational effort. However, quantum corrections
considerably reduce the naive gain factor. Also
there is some overhead from the nonperturbative matching of fermionic
and gluonic anisotropies, which is particularly delicate when sea
quarks are included. Nevertheless, a pilot study with $n_f=2$
and $\xi\approx 6$ already exists.\cite{Juge:2006kk}

We summarize the quenched situation in Fig.~\ref{fig:chquench}.
The Columbia\cite{Liao:2002rj} and CP-PACS\cite{Okamoto:2001jb}
results are extrapolated to the continuum limit. Similar
results for a smaller number of states have been obtained
by the Guangzhou group\cite{Mei:2002ip,Luo:2005zg} and 
QCD-TARO.\cite{Choe:2003wx} It is reassuring that the latter
study on isotropic lattices confirms the anisotropic results.
The scale $r_0^{-1}\approx 394$~MeV
is set from potential models. A systematic scale error of
about 10 \%
should be assumed, for stable states that do not mix!
One often thinks of the charm quark
as heavy in the sense that its mass is bigger than typical
QCD binding energies.
However, the spectrum of states entirely made out of glue
covers a similar energy range. To illustrate this
we have superimposed these glueballs\cite{Chen:2005mg} onto the Figure.

Note that for the lattice results displayed, effects of
diagrams with disconnected quark lines have been neglected. One might expect
OZI violating contributions from these, for instance for states that lie close
to glueballs with the same quantum numbers or close to decay thresholds.
Also (tiny) effects from the axial anomaly or from $\eta_c-\eta'$ mixing
cannot be excluded. The impact of such diagrams on the vector
and pseudoscalar channels has been studied
by McNeile and Michael\cite{McNeile:2004wu} and by
CP-PACS\cite{deForcrand:2004ia} and found to be insignificant
within errors of about 20~MeV.

The Figure also contains the lightest two
spin-exotic $c\bar{c}$-gluon
hybrids.\cite{Liao:2002rj} Again, the results agree with
other studies.\cite{Bernard:1997qe,Mei:2002ip,Luo:2005zg}
It is also possible to employ hybrid sources with
non-exotic quantum numbers.
States prepared in this way are found to decay rapidly in Euclidean time
into the respective non-hybrid ground
states\cite{Liao:2002rj,Mei:2002ip,Luo:2005zg}
which is hardly surprising, given the fact that one
expects an energy gap of about
1.4~GeV when solving the Schr\"odinger equation with ground state
and hybrid lattice potentials as input. However, 
the extent of spectroscopic hybrid components in
charmonium wavefunctions deserves further
study.\cite{Burch:2001tr,Luo:2005zg}
Similarly, it is difficult to disentangle
$S$ and $D$ waves in lattice simulations of the
$1^{--}$ sector.\cite{Liao:2002rj}

When sea quarks are switched on, would-be glueballs, 
standard charmonia and $D\overline{D}$ molecules
will mix (as will exotic hybrid charmonia and molecules).
This has not yet been studied on the lattice.
In addition, one would expect the fine structure
to be particularly sensitive to the sea quark content.
For instance, the $S$ wave singlet-triplet splitting
in NRQCD reads,
$\Delta M=c_F/(6m_c^2)\langle\psi|V_4|\psi\rangle+\cdots$ where
$c_FV_4=(32\pi/3)\alpha_s\delta^3({\mathbf r})$ to leading order
in perturbative QCD. Due to the short-distance nature of
this term one would expect
significant finite $a$ effects. Moreover, with sea quarks
included, the running of $\alpha_s(q)$ will slow down such
that at high momenta/short distances $\alpha_s$ will maintain a
relatively larger value.
In Ref.\cite{Bali:1998pi} this effect is estimated to amount
to 30--40~\%. 

Indeed, setting the scale from $r_0$, a splitting of only
$77(2)(6)$ MeV is found.\cite{Choe:2003wx} When adjusting the lattice
spacing to the $1\overline{P}-1\overline{S}$ splitting instead,
the central value
increases to $\Delta M\approx 89$~MeV,
still short of the experimental $\Delta M\approx 117$~MeV.
This result is consistent with Refs.\cite{Liao:2002rj,Okamoto:2001jb}
while in small volume studies at one lattice spacing\cite{Tamhankar:2005zi}
and at a too small charm quark mass\cite{Dudek:2006ej} higher results
have been obtained. The FNAL and MILC\cite{Gottlieb:2005me} collaborations
indeed find the $J/\psi-\eta_c$ splitting
to increase to $\Delta M =94(1), 101(2)$ and
$107(3)$ MeV at lattice spacings
$a^{-1}\approx 1.1, 1.6$ and 2.3~GeV, respectively,
in studies with\cite{xxx} $n_f\stackrel{?}{=}2+1$
sea quark flavours.
Note however that there are theoretical
doubts about the validity of MILC's fermion approach. Also, disconnected quark
line diagrams have not yet been included. While their impact is
likely to be small effects of up to 20~MeV cannot be
excluded.\cite{McNeile:2004wu}

\section{Radiative Transitions}
In order to match experimental
resonances to theoretical predictions ultimately one has to
go beyond the mass spectrum. For instance
one would expect hybrids and molecules to couple 
differently to decay and production channels than states with
quark model dominance. Gaining information on strong decay rates from lattice
simulations, without additional assumptions, is complicated.
However, calculating electromagnetic (EM) decay and transition
rates is in principle straight-forward. For narrow states
such predictions can directly be confronted
with experiment. For broad states above threshold,
where data on EM transitions is lacking,
a knowledge of the wavefunction will reveal qualitative
information on possible production mechanisms.

\begin{figure}[t]
\centerline{\includegraphics[width=0.8\textwidth, bb=14 43 710 522, clip]{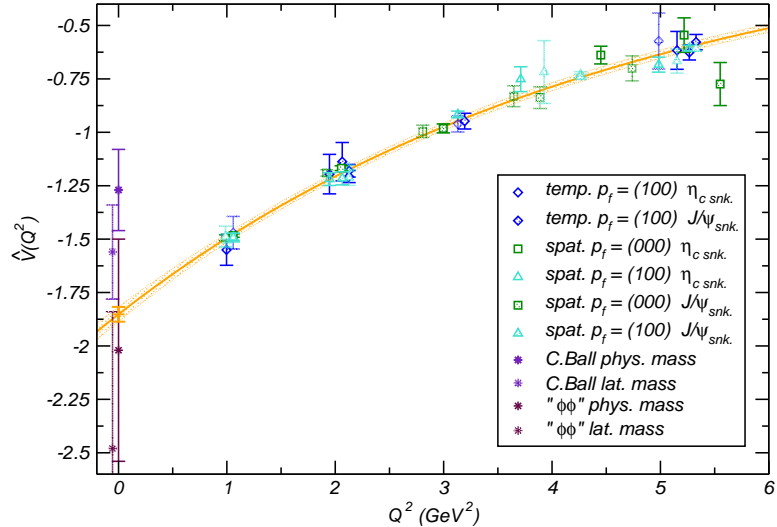}}
\caption{The $J/\psi\rightarrow\eta_c\gamma_{M1}$ matrix element
$\hat{V}(Q^2)$, together with $\hat{V}(0)$, extracted from
experiment.\protect\cite{Dudek:2006ej}}
\label{fig:dudek}
\end{figure}

QCD-TARO\cite{Choe:2003wx} calculated electric quarkonium wavefunctions.
In the non-relativistic approximation, the EM decay constant
is to leading order proportional
to the wavefunction at the origin. 
Dudek and Edwards directly calculated two-photon
decay rates\cite{Dudek} as well as the decay
constants\cite{Dudek:2006ej} $f_{J/\psi}, f_{\psi'},
f_{\eta_c}$ and $f_{\eta_c'}$. These studies have been performed
at one lattice spacing on small volumes in the quenched approximation
and with a charm quark mass about 5~\% smaller than the physical one.
In particular the calculations of this pilot study of EM transition
rates are extremely promising. For instance the M1
transition rate $\Gamma(J/\psi\rightarrow \eta_c\gamma)\propto
{\mathbf q}^3/(m_{J/psi}+m_{\eta_c})^2\alpha_{{\mbox{\tiny fs}}}|\hat{V}(0)|^2$
contains the term $\hat{V}(Q^2)$ at $Q=0$.
This decomposes into a kinematical factor,
multiplied by the
three point function, $\langle\eta_c({\mathbf p}')|j^{\mu}({\mathbf 0})|
J/\psi({\mathbf p})\rangle$, that can be calculated on the lattice.
In Fig.~\ref{fig:dudek} we display the lattice prediction\cite{Dudek:2006ej}
of $\hat{V}$, obtained for different momentum
transfers $Q^2=({\mathbf p}'-
{\mathbf p})^2$, together with the $Q=0$ value derived from the
experimentally measured transition rate. This expectation was
phenomenologically corrected for the mis-adjusted lattice charm
quark mass (lat.\ mass) to facilitate the comparison: the lattice
$Q^2$-extrapolation appears very reasonable.

The transition $\chi_{c1}\rightarrow J/\psi\gamma$ was studied too
as were $\chi_{c0}\rightarrow J/\psi\gamma$ and
$\Gamma(h_c\rightarrow\eta_c\gamma)\approx 600$~keV. The latter
rate is a prediction since CLEO only detected the cascade
$\psi'\rightarrow \pi_0 h_c, h_c\rightarrow \eta_c\gamma$ where
both individual rates are unknown.

\section{Mixing and Strong Decays}
It is evident that states close to decay thresholds
like the $X(3872)$ or the $Y(3940)$ (see Fig.~\ref{fig:spec})
will couple to molecular $D\overline{D}$ or
$D_s\overline{D}_s$ components, respectively. At present
there exist no meaningful
direct lattice studies of such $c\bar{c}\leftrightarrow
c\bar{q}q\bar{c}$ mixing phenomena. Ideally one would
create and destroy states using an extended basis of operators, including
molecules and hybrids, to study their mixing. 

\begin{figure}[th]
\centerline{\includegraphics[width=0.45\textwidth]{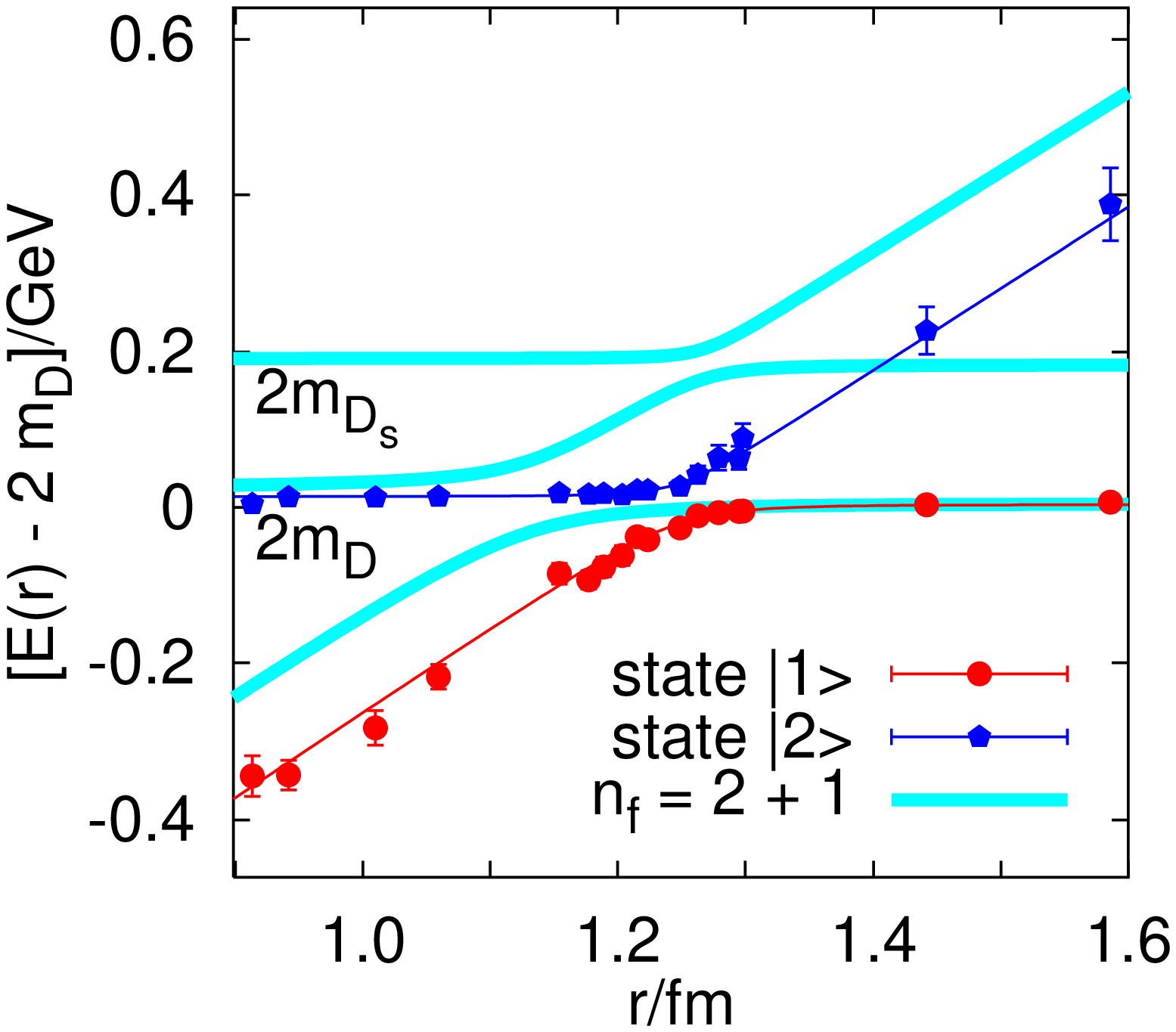}~
\includegraphics[width=0.5\textwidth]{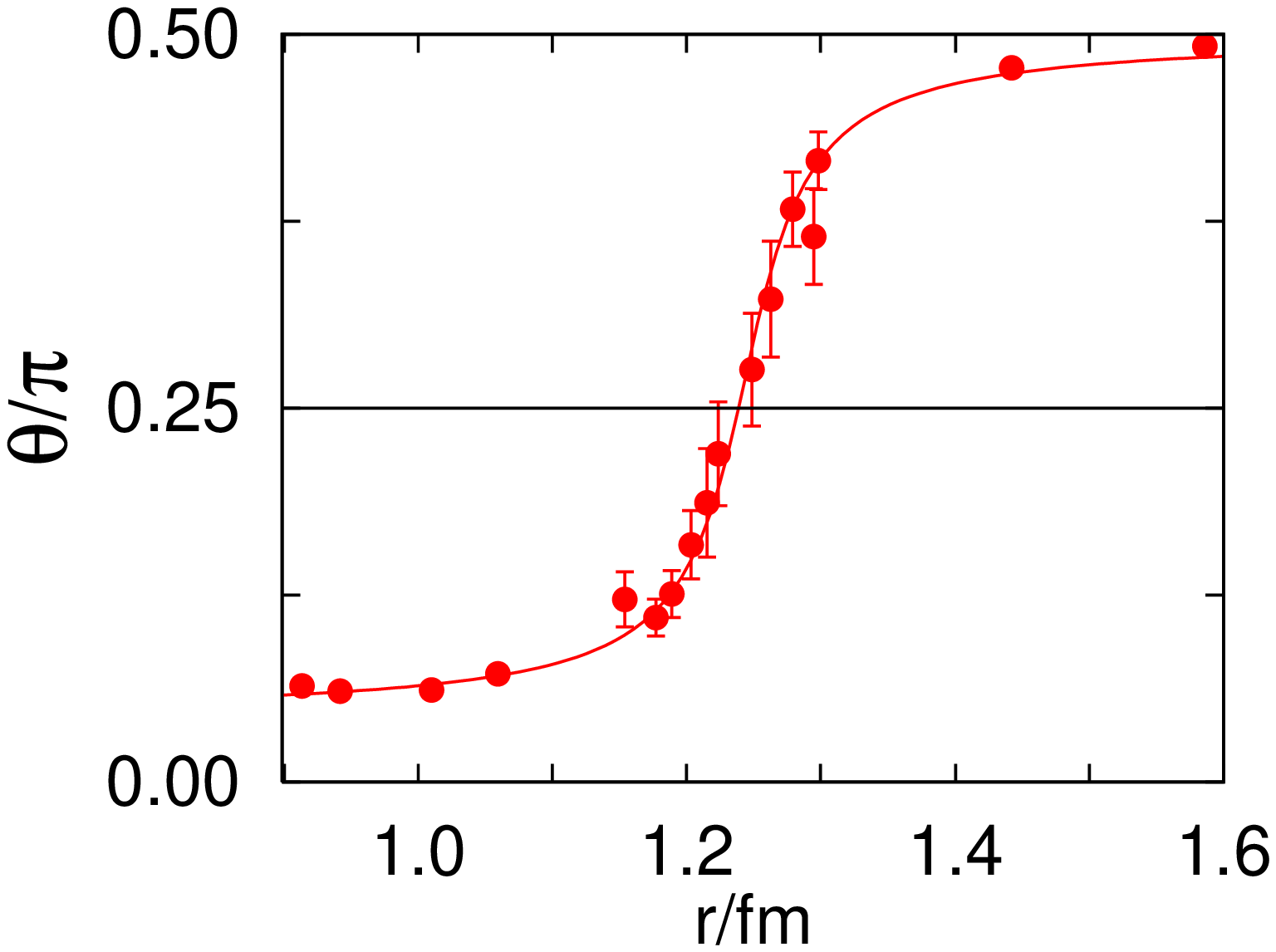}}
\caption{The $Q\overline{Q}|{\mathcal D}\overline{\mathcal D}$ energy levels
and mixing angle for $n_f=2$.\protect\cite{Bali:2005fu}}
\label{fig:mixing}
\end{figure}
In the absence of such an investigation, one might
benefit from the non-relativistic
approximation. At least qualitatively charmonia can
be described by potential models. The $Q\overline{Q}$
sector can be extended to include ${\mathcal D}\overline{\mathcal D}$
potentials.
$Q$ denotes a static (heavy) quark and
${\mathcal D}=Q\bar{q}$. This situation has recently been studied on the
lattice,\cite{Bali:2005fu} albeit only at one lattice spacing
$a\approx 0.083$~fm and with $n_f=2$ mass-degenerate sea quark flavours,
slightly lighter than the strange quark. We denote the
physical eigenstates by $|n\rangle$. These can be decomposed,
\begin{eqnarray}
\label{eq:1}
|1\rangle&=&\cos{\theta}\,|Q\overline{Q}\rangle+\sin{\theta}
\,|{\mathcal D}\overline{\mathcal D}\rangle+\cdots,
\\
|2\rangle&=&-\sin{\theta}\,|Q\overline{Q}\rangle+\cos{\theta}
\,|{\mathcal D}\overline{\mathcal D}\rangle+\cdots.
\label{eq:2}
\end{eqnarray}
The resulting energy levels and mixing angle $\theta$ are displayed 
in Fig.~\ref{fig:mixing} as a function of the $Q\overline{Q}$-separation
$r$.

At $r>r_c\approx 1.25$~fm the
$Q\overline{Q}$ string breaks and the ground state is
dominated by the ${\mathcal D}\overline{\mathcal D}$
configuration. Interestingly,
also at $r\ll r_c$ there is a significant four-quark admixture in
the ground state. This 4~\% probability is consistent with
the large four-quark components found in phenomenological
models of charmonia and bottomonia.\cite{Eichten:2004uh} 
In Ref.\cite{Bali:2005bg} the action and energy density distributions
around the static sources have been investigated:
string breaking appears to be an instantaneous process with
no spatial localization of the light $q\bar{q}$ pair that is created.
This suggests an $2\times 2$ Hamiltonian with a transition term
$(E_2-E_1)\sin\theta\cos\theta$ 
that only depends on the distance $r$.

One can include
${\mathcal D}\overline{\mathcal D}$ effects
as quantum mechanical
perturbations,\cite{Eichten:2004uh,Drummond:1998eh}
starting from an unperturbed $Q\overline{Q}$
wave function $\psi_0$.
By summing over all possible intermediate states $i$
where $\psi_i$ can be in both sectors,
${\mathcal D}\overline{\mathcal D}$ and $Q\overline{Q}$,
one ends up with complex Hamiltonian for each state.
The solution of the corresponding
Schr\"odinger problem yields both: the respective
energy levels and decay widths.

A study of the sea quark mass dependence of the static energy levels
is as yet lacking. The bands in the left plot
of Fig.~\ref{fig:mixing} represent
an $n_f=2+1$ speculation. In this case string breaking
will occur at a somewhat smaller distance, $r_c=1.13(10)(10)$~fm,
the gap $E_2(r_c)-E_1(r_c)$ will be larger and moreover
there will be a second string breaking threshold
$Q\overline{Q}\leftrightarrow{\mathcal D}_s\overline{\mathcal D}_s$.
It would be interesting to see, once the sea quark mass
dependence has been established, if properties
of $X(3872)$ and $Y(3940)$ can be reproduced or if the
present overpopulation of the $1^{--}$ sector can be explained
by coupled channel models, based on such lattice potentials.

\section{Summary}
We have arrived at a good understanding of the spectrum of states
in the mass region, 2.5 -- 5 GeV, in the
quenched approximation to QCD, including glueballs and $c\bar{c}$-glue
hybrid states, besides standard charmonia. However,
information on four-quark states is still lacking and there is
some uncertainty regarding the size of contributions due to
diagrams with disconnected quark lines.
Electromagnetic decay and
transition rates have been calculated successfully in a pilot study.

Studies of $n_f=0$ QCD set the stage for more realistic
$n_f>0$ simulations. Promising $n_f\stackrel{?}{=}2+1$ results have
been obtained, indicating a convergence of low lying states
towards experiment, once sea quarks are included. Over the next
few years such simulations will be extended to include more
realistic sea quarks and a broader basis of operators to create states,
including four-quark and hybrid operators. In the meantime
one has to resort to coupled channel models to understand
threshold effects. These can be refined by using input from
lattice simulations. Ultimately, one would wish to arrive at
QCD predictions of strong decay rates. This is very hard to
achieve by a direct lattice computation but again,
the predictive power of models can benefit from dedicated
lattice tests and comparisons.
Last but not least I hope that these exciting times 
will continue with further experimental discoveries.

\section*{Acknowledgements}
I thank the organisers of the conference for the invitation.
I gratefully acknowledge support from the
EC Hadron Physics I3 Contract RII3-CT-2004-506078.

\end{document}